\documentclass[prd,aps,twocolumn,nofootinbib,superscriptaddress,eqsecnum,floatfix,preprintnumbers,amsmath,amssymb,longbibliography]{revtex4-2}

\makeatletter
\renewcommand{\p@subsection}{}
\renewcommand{\p@subsubsection}{}
\makeatother

\usepackage{amsmath}
\makeatletter
\@removefromreset{equation}{section}

\makeatother

\usepackage{graphicx}
\usepackage{epsfig}
\usepackage[dvipsnames]{xcolor}
\usepackage[utf8]{inputenc}
\usepackage{stmaryrd}
\usepackage{mathrsfs}
\usepackage{mathalfa}
\usepackage{accents}
\usepackage{enumitem}
\usepackage[normalem]{ulem}

\usepackage[colorlinks=true,
            citecolor=green,
            linkcolor=red,
            filecolor=cyan,
            urlcolor=magenta,
            backref=false]{hyperref}

\newcommand{\be}{\begin{equation}}
\newcommand{\ee}{\end{equation}}
\newcommand{\bea}{\begin{eqnarray}}
\newcommand{\eea}{\end{eqnarray}}
\newcommand{\ba}{\begin{equation}\begin{aligned}}
\newcommand{\ea}{\end{aligned}\end{equation}}
\newcommand{\beg}{\begin{gather*}}
\newcommand{\eng}{\end{gather*}}
\newcommand{\hh}{,\hspace{0.5cm}}
\newcommand{\hhh}{,\hspace{0.2cm}}

\newcommand{\n}[1]{\label{#1}}

\newcommand{\CAL}{\mathcal}

\newcommand{\bs}{\begin{split}}
\newcommand{\es}{\end{split}}

\begin{document}

\title{Modified Double Copy for Quasitopological Gravity with Matter}

\author{Valeri P. Frolov }
\email[]{vfrolov@ualberta.ca}
\affiliation{Theoretical Physics Institute, Department of Physics,
University of Alberta,\\
Edmonton, Alberta, T6G 2E1, Canada
}


\begin{abstract}
We extend the modified double-copy construction to quasitopological gravity (QTG) coupled to matter. Spherically symmetric QTG solutions in $D$ dimensions are generated from an auxiliary gauge field in a flat $(D+1)$-dimensional spacetime. The gravitational field equations reduce to two relations determined by the QTG generating function and the matter stress--energy tensor. For a broad class of spherically symmetric matter sources, solutions of the auxiliary gauge-field equations can be used to construct exact solutions of the QTG field equations. The framework applies to Maxwell theory, nonlinear electrodynamics, and Yang--Mills theory, encompasses both static and Vaidya-type solutions, and reproduces the Einstein-gravity limit.
\end{abstract}
\hfill    Alberta Thy 3-26

\maketitle

\section{Introduction}

The classical double-copy correspondence establishes a remarkable connection between gauge theory and gravity in $D$ dimensions by relating solutions of Maxwell theory in a flat $D$-dimensional spacetime with metric $ds_0^2$ to Kerr--Schild solutions of the Einstein equations \cite{KerrSchild1965}. The latter take the form
\be\n{KKSS}
ds^2=ds_0^2+\Phi\,(k_{\mu}dx^{\mu})^2\, ,
\ee
where, $k_{\mu}$ is a null eigenvector of the Maxwell field and $\Phi$ is its potential (for review see e,g,
 \cite{Visser:2007fj} and references therein). This method can be applied to wide class of solutions of the Einstein equations \cite{Monteiro_2014,Luna_2015,White_2018,bern2019duality}.

Recently, this correspondence was extended in Ref.~\cite{Frolov:2025ddw} to a modified double-copy construction formulated in an auxiliary $(D+1)$-dimensional flat spacetime, opening the possibility of applying double-copy ideas beyond Einstein gravity.

In this paper, we demonstrate that the modified double-copy formalism provides a natural framework for constructing exact solutions of quasitopological gravity (QTG) coupled to matter. QTG is a higher-curvature theory that admits regular black-hole solutions while retaining many of the remarkable properties of Einstein gravity
\cite{Bueno:2019ycr,Bueno:2019ltp, Bueno:2024dgm, Frolov:2024lza, Bueno:2025collapse, Bueno:2025formation, Bueno:2026singularmatter, Frolov:2026massinflation, Bueno:2026nonlocal, Borissova:2026rbi,Borissova:2026prd, Borissova:2026wmn}. We show that the QTG field equations in a curved spacetime $\CAL{M}^D$ can be reduced to a set of equations for an auxiliary gauge field in a flat $(D+1)$-dimensional spacetime. In the Einstein limit, these equations reduce to the linear Maxwell equations, whereas for a generic QTG model they describe a nonlinear electrodynamics uniquely determined by the generating function that specifies the theory.

The formalism applies to a broad class of matter sources, including Maxwell, nonlinear-electrodynamics, and Yang--Mills fields. It provides a unified procedure for generating both static and Vaidya-type solutions, with Einstein gravity recovered as a special case.

\section{QTG Equations}
\label{Sec2}

Quasitopological gravity (QTG) extends the Einstein--Hilbert action by
an infinite series of higher-curvature terms,
\begin{equation}
\label{QTA}
S_{\rm QTG}
 =\frac{1}{2\varkappa}\int d^Dx\,\sqrt{-g}
 \left[R+\sum_{j\geq 2}\alpha_j\ell^{2(j-1)}Z_j\right].
\end{equation}
Here $\varkappa=8\pi G^{D}$ and $G^D$ is the $D-$dimensional gravitational coupling constant.
$Z_j$ denotes a linear combination of scalar invariants, each formed by contracting a product of $j$ Riemann tensors.
The invariants $Z_j$ are chosen so that the field equations remain
second order under spherical symmetry \cite{Bueno:2019ycr}. Their
couplings are conveniently encoded in the generating function
\begin{equation}
h(p)=\sum_{j\geq 1}\alpha_j \ell^{2(j-1)} p^j,
\qquad \alpha_1=1,
\end{equation}
The parameter $\ell$ has dimension of the {\em length}. It determines the scale at which the higher curvature contribution becomes important.

When $h(p)=p$, the theory reduces to Einstein gravity.
For a spherically symmetric spacetime,
\begin{equation}
\label{GO}
ds^2=\gamma_{\mu\nu}(x)dx^\mu dx^\nu
     +r^2(x)d\Omega_{D-2}^2, \, \ \mu,\nu=0,1,
\end{equation}
where $d\Omega_{D-2}^2$ denotes the metric on the unit $(D-2)$-sphere,
the quantity
\begin{equation}
p=\frac{1-(\nabla r)^2}{r^2}
\end{equation}
is the primary curvature invariant \cite{Frolov:2024lza}.
We assume that $h(p)$ is analytic and invertible on the physical branch.

Let us note that any spherically symmetric tensor $P_{AB}$
 $(A,B=0,1,\ldots , D-1)$
can be decomposed as
\begin{equation}
\label{PPPP}
P_A{}^B=\frac{1}{r^{D-2}}
\left[
 \delta_A^\mu\delta_\nu^B\,\mathcal{P}_\mu{}^\nu
 +\delta_A^i\delta_j^B\,
  \frac{\mathcal{P}}{D-2}\delta_i^j
\right].
\end{equation}
Its conservation law, $\nabla_B P^B{}_A=0$, reduces to
\begin{equation}
\label{G_CONS}
\mathcal{P}_\mu{}^{\nu}{}_{;\nu}
 =\frac{r_{;\mu}}{r}\,\mathcal{P}.
\end{equation}
Here and throughout, a semicolon denotes covariant differentiation with respect to the two-dimensional metric $\gamma_{\mu\nu}$.

Using this tensor decomposition, the equations of
the QTG model coupled with matter can be written the form
\be \n{GEQ}
\CAL{G}^{\mu\nu}=\frac{2\kappa}{D-2}\CAL{T}^{\mu\nu}\hh \CAL{G}=\frac{2\kappa}{D-2}\CAL{T}\, .
\ee

We use ingoing null coordinates $(v,r)$, in which\footnote{All the results remain valid in outgoing null coordinates upon replacing $v\to u$, where $u$ denotes the outgoing (retarded) time.}
\begin{equation}
\label{NFR}
ds^2=-N^2 fdv^2    +2Ndv\,dr+r^2d\Omega_{D-2}^2 ,
\end{equation}
where $N$ and $f$ are functions of $v$ and $r$.
For this metric, the independent orbit-space components of the QTG
field equations \cite{Frolov:2026tft}
\begin{equation}\label{EQQ}
\begin{split}
&\mathcal{G}_{vv}+Nf\mathcal{G}_{vr}=N H_{,v}\hh
\mathcal{G}_{vr}=-N H_{,r},\\
&\mathcal{G}_{rr}=2r^{D-3}\frac{N_{,r}}{N}h'(p)\hh  H=r^{D-1}h(p).
\end{split}
\end{equation}

We consider matter sources whose reduced stress-energy tensor has the
form
\begin{equation}
\label{SETT}
\mathcal{T}_{\mu\nu}
 =\tau(v,r)\gamma_{\mu\nu}
  +\sigma(v,r)k_\mu k_\nu,
\qquad
k_\mu=v_{,\mu}.
\end{equation}
The angular component is denoted by $\mathcal{T}$ according to Eq.~\eqref{PPPP}. The term $\sigma(v,r)$ describes the energy flux of null matter. In its absence, the QTG field equations imply the existence of a Killing vector, and hence a generalized Birkhoff theorem holds \cite{Frolov:2026tft}.

Since $\mathcal{T}_{rr}=0$, the $rr$ field equation
implies $N_{,r}=0$. A redefinition of $v$ therefore allows one to choose
\begin{equation}
N=1\, ,
\end{equation}
and the metric \eqref{NFR} then assumes the Kerr--Schild form \eqref{KKSS} with
flat metric
\be \n{MET0}
ds_0^2=-dv^2+2dv dr +r^2 d\Omega_{D-2}^2\, ,
\ee
and $\Phi=r^2 p$.
The conservation law \eqref{G_CONS} gives
\begin{equation}
\sigma_{,r}=-\tau_{,v},
\qquad
\tau_{,r}=\frac{\mathcal{T}}{r}.
\label{EEQQ}
\end{equation}
Remarkably, these equations are independent of $f$ and therefore hold
both in the full curved spacetime $\CAL{M}^D$ and in the flat background
$M^D$. The remaining QTG equations reduce to
\begin{equation}
\label{HHVR0}
H_{,v}=\frac{2\kappa}{D-2}\sigma,
\qquad
H_{,r}=-\frac{2\kappa}{D-2}\tau .
\end{equation}
The first relation in \eqref{EEQQ} is precisely the integrability
condition for this system while the second relation allows one to find $\CAL{T}$ for known $\tau$. Suppose the function $H$ is given at some point $(v_0,r_0)$ of the $(v,r)$ plane, Then its value at $(v,r)$ can be found as an integral  $\int H_{\mu}dx^{\mu}$ along a line connecting these points. The integral does not depend on the chosen path.
Thus, once the conserved matter distribution
$(\tau,\sigma,\mathcal{T})$ in the physical spacetime $\CAL{M}^D$ is specified, Eq.~\eqref{HHVR0} determines
$H(v,r)$. The metric $ds^2$ is then reconstructed from
\begin{equation}
\begin{split}
&h(p)=\frac{H(v,r)}{r^{D-1}}\hhh
p=h^{-1}\!\left(\frac{H(v,r)}{r^{D-1}}\right)\, ,\\
&f(v,r)=1-r^2p(v,r).
\end{split}
\end{equation}
This reduction contains all information required for the modified
double-copy construction developed below.

\section{Modified double-copy method of solving QTG equations}
\label{Sec3}

The modified double-copy formalism was proposed in Ref.~\cite{Frolov:2025ddw} where it was used to construct vacuum spherically symmetric solutions of the QTG equations.
We now demonstrate that it can be applied to QTG equations in a curved spacetime $\CAL{M}^D$ in the presence of matter.
The construction starts with an auxiliary nonlinear electromagnetic field defined in a flat $(D+1)$-dimensional spacetime $M^{D+1}$. Its restriction to a $D$-dimensional hyperplane is then used to reconstruct the Kerr--Schild metric on $\mathcal{M}^D$.\footnote{Let us emphasize that this electromagnetic field plays a purely auxiliary role. In QTG coupled to nonlinear electrodynamics, it should not be confused with the physical electromagnetic field that acts as a source of the gravitational field.}

Denote by $X^a=(T,X_1,\ldots,X_{D-1},X_D)$  Cartesian coordinates in Minkowski space
$M^{D+1}$ with metric $dS_0^2$. We define its "equatorial hyperplane" $\Pi$ by condition $X_D=0$.
This hyperplane is a $D-$ dimensional Minkowski space  with the flat metric $ds_0^2$.
Denote
\be
R^2=\sum_{a=1}^{D} X_a^2\hhh
V=T+R \hhh K_a=V_{,a} \, .
\ee
Then on the equatorial hyperplane $\Pi$ the metric $dS_0^2$ coincides with \eqref{MET0} and
one has
\be
R\big|_{\Pi}=r\hh
V\big|_{\Pi}=v\hh
K_A\big|_{\Pi}=k_A .
\ee
Here $\big|_{\Pi}$ denotes restriction to $\Pi$.

Consider nonlinear electrodynamics in $M^{D+1}$ with action
\be\n{NLM}
\begin{split}
W={}&\int d^{D+1}X\,L(\CAL{E})
+\int d^{D+1}X\,\CAL{J}^a\CAL{A}_a,\\
&\CAL{E}^2=-\frac12\CAL{F}_{ab}\CAL{F}^{ab}.
\end{split}
\ee
We assume the weak-field behavior $L(\CAL{E})=\frac12\CAL{E}^2+O(\CAL{E}^3)$.
For a spherically symmetric configuration, the auxiliary field $\CAL{A}_a$ and
its current $\CAL{J}^a$ have only $(V,R)$ components, and
\be
\CAL{F}_{VR}=\CAL{A}_{V,R}-\CAL{A}_{R,V}\hh
\CAL{E}=\frac12 e^{\mu\nu}\CAL{F}_{\mu\nu}
       =\CAL{F}_{VR},
\ee
where $e^{\mu\nu}$ is the unit antisymmetric tensor in the
two-dimensional $(V,R)$ sector.

After integration of the action \eqref{NLM} over the angular variables and division by
$\Omega_{D-1}$, the reduced action becomes
\be\n{WWAA}
\CAL{W}=\int d^2X\,R^{D-1}
\left[L(\CAL{E})+\CAL{J}^{\mu}\CAL{A}_{\mu}\right].
\ee
To establish the correspondence between the auxiliary-field action and the QTG model, we choose the Lagrangian $L(\CAL{E})$ such that
\be \n{LLhh}
\frac{dL}{d\CAL{E}}=h(\CAL{E})\, ,
\ee
where the function $h$ specifies the QTG model.
Variation of the action \eqref{WWAA}
with respect to $\CAL{A}_\mu$ gives equations
\be\n{NELEQ}
\begin{split}
\CAL{H}_{,R}&=-R^{D-1}\CAL{J}^{V},\\
\CAL{H}_{,V}&=\phantom{-}R^{D-1}\CAL{J}^{R},
\end{split}
\ee
where
\be
\CAL{H}=R^{D-1}h(\CAL{E})\, .
\ee
Their compatibility condition is
\be\n{CONS}
\partial_V j^{V}
+\partial_R j^{R}=0,
\ee
which is the conservation law for the reduced current
$j^\mu=R^{D-1}\CAL{J}^\mu$.

To connect this system with QTG, note that the restriction of any
spherically symmetric scalar $B(V,R)=B(T+R,R)$ to $\Pi$ satisfies
\be
B(V,R)\big|_{\Pi}=B(v,r).
\ee
Thus, such fields in $M^{D+1}$ are in one-to-one correspondence with
their restrictions to $M^D$. We choose the auxiliary current as
\be\n{CURR}
\begin{split}
\CAL{J}^{V}
 &=\frac{2\kappa}{D-2}\,
   \frac{\tau(V,R)}{R^{D-1}},\\
\CAL{J}^{R}
 &=\frac{2\kappa}{D-2}\,
   \frac{\sigma(V,R)}{R^{D-1}},
\end{split}
\ee
where $\tau(V,R)$ and $\sigma(V,R)$ are smooth extensions of stress-energy components
$\tau(v,r)$ and $\sigma(v,r)$ away from $\Pi$. Equation~\eqref{CONS}
then becomes
\be\n{CONS1}
\tau_{,V}+\sigma_{,R}=0,
\ee
whereas Eqs.~\eqref{NELEQ}, restricted to $\Pi$, give
\be\n{HHVR1}
H_{,v}=\frac{2\kappa}{D-2}\sigma
\hh
H_{,r}=-\frac{2\kappa}{D-2}\tau,
\ee
with
\[
H(v,r)=\CAL{H}(V,R)\big|_{\Pi}.
\]
These are precisely the QTG equations \eqref{HHVR0}.

The modified double-copy construction can therefore be summarized as follows. For a given
QTG model specified by $h(p)$, one chooses the auxiliary nonlinear
electrodynamics Lagrangian so that
\be
\frac{dL(\CAL{E})}{d\CAL{E}}=h(\CAL{E}).
\ee
After solving Eqs.~\eqref{NELEQ} with the current \eqref{CURR}, one
restricts the solution to $\Pi$ and identifies
\be
\CAL{E}\big|_{\Pi}=p\hh
\big(\CAL{H}/R^{D-1}\big)\big|_{\Pi}=h(p).
\ee
Inverting $h=h(p)$ determines $p(v,r)$, and the metric function follows
from
\be\n{ffpp}
f(v,r)=1-r^2p(v,r).
\ee
Together with the gauge choice $N=1$, this yields the required solution of the QTG equations with the matter source \eqref{PPPP}--\eqref{SETT}. For $h(p)=p$, the auxiliary gauge field obeys the linear Maxwell equations, and the corresponding double-copy construction reduces to Einstein gravity.

Let us emphasize an important distinction between the modified double-copy approach and the standard double-copy formalism. In the modified construction, the metric function \eqref{ffpp} is reconstructed by identifying the gravitational curvature invariant $p$ of the curved spacetime $\CAL{M}^D$ with the auxiliary electrostatic field invariant $\CAL{E}$ in a flat spacetime $M^{D+1}$. It is precisely this identification that allows the double-copy construction to be extended from Einstein gravity to QTG.

Up to this point, the source of the gravitational field has been treated as a prescribed matter distribution. However, the same stress--energy structure arises naturally in Maxwell theory, nonlinear electrodynamics, and Yang--Mills theory.
This allows one to use the modified double-copy formalism for study regular charged black hole solutions for QTG coupled with Maxwell and nonlinear electrodynamica \cite{Hennigar:2025yqm,Hao:2025utc,PinedoSoto:2026hfm}, as well as their Vaidya type generalizations \cite{Frolov:2026tft}.
This demonstrates that the modified double-copy construction developed here applies to a broad class of models in which QTG is coupled to physical matter fields.

\subsection{Example}

As a simple illustration of the map to $M^{D+1}$, consider a physical Maxwell field in $\CAL{M}^D$  with action
\be
S_{\rm m}=-(16\pi)^{-1}\int d^Dx\sqrt{-g}\,F_{AB}F^{AB}\, ,
\ee
For a static, spherically symmetric electric field,
$F_{vr}=q/r^{D-2}$, where $q$ is a constant charge parameter.
Its reduced stress tensor, in the conventions of
Eqs.~\eqref{PPPP} and \eqref{SETT}, is
\begin{equation}
\label{ex:Maxwell-source}
\tau=-\frac{q^2}{8\pi r^{D-2}},\qquad
\sigma=0,\qquad
\mathcal{T}=\frac{(D-2)q^2}{8\pi r^{D-2}}.
\end{equation}
Thus, the auxiliary current prescribed by Eq.~\eqref{CURR} is
\begin{equation}
\label{ex:aux-current}
\begin{split}
\mathcal{J}^{V}&=-\frac{(D-3)Q^2}{R^{2D-3}},\qquad
\mathcal{J}^{R}=0,\\
Q^2&=\frac{\kappa q^2}{4\pi(D-2)(D-3)}.
\end{split}
\end{equation}
Integrating Eq.~\eqref{NELEQ} gives, for $R>0$,
\begin{equation}
\label{ex:aux-solution}
\begin{split}
\mathcal{H}(R)&=\mu-\frac{Q^2}{R^{D-3}},\\
\mathcal{E}(R)&=h^{-1}\!\left(
\frac{\mu}{R^{D-1}}-\frac{Q^2}{R^{2D-4}}\right),
\end{split}
\end{equation}
where $\mu$ is an integration constant. Restriction to $\Pi$ and
the identification $p=\mathcal{E}|_\Pi$ therefore yield the following expression for the metric function
\begin{equation}
\label{ex:charged-map}
f(r)=1-r^2h^{-1}\!\left(
\frac{\mu}{r^{D-1}}-\frac{Q^2}{r^{2D-4}}\right).
\end{equation}

\section{Discussion}\label{Sec5}

We have extended the modified double-copy formalism to
quasitopological gravity coupled to matter. The construction replaces
the direct solution of the nonlinear QTG field equations by an
auxiliary nonlinear electrodynamics problem in a flat
$(D+1)$-dimensional spacetime, from which the corresponding
Kerr--Schild metric is reconstructed.

The auxiliary gauge field $\CAL{A}_a$ may be interpreted as a single-copy field. This construction differs from the standard double-copy formalism, in which the single-copy gauge field is defined in $D$ dimensions, matching the dimension of the gravitational spacetime $\CAL{M}^D$, rather than in $(D+1)$ dimensions.
An action for a corresponding $D$-dimensional gauge field $A_B$ ($B=0,1,\ldots ,D-1)$) can be obtained from \eqref{WWAA}. To this end, we define $E=A_{v,r}-A_{r,v}$ and set $E=rp$. The action \eqref{WWAA} then reduces to
\be\n{DWWAA}
\CAL{W}^D=\int d^2X,r^{D-2}
\left[\CAL{L}(E,r)+\CAL{J}^{\mu}A_{\mu}\right]\, ,
\ee
where $J^{\mu}=r\CAL{J}^{\mu}\big|_{\Pi}$ and
\be
\CAL{L}(E,r)=r^2 L(E/r)\, .
\ee
Varying this action with respect to $A_v$ and $A_r$ reproduces equations \eqref{HHVR0}.

For Einstein gravity, $h(p)=p$, the effective $D$-dimensional Lagrangian $\CAL{L}$ is proportional to $E^2$, and the auxiliary field $A_B$ satisfies the standard $D$-dimensional Maxwell equations. Thus, this construction recovers the standard single copy of the gravitational field. For a general QTG theory, however, $h(p)$ is nonlinear, and $\CAL{L}$ depends explicitly on $r$. Consequently, this reduced action does not, in general, arise from the spherical reduction of a $D$-dimensional covariant gauge theory built solely from the metric and the electromagnetic field strength.

For spherically symmetric configurations, the entire dependence on the
underlying gravity theory is encoded in a single generating function $h(p)$,
whereas the matter sector enters only through its stress--energy
tensor. The formalism applies to Maxwell theory, nonlinear
electrodynamics, and Yang--Mills theory, providing a unified treatment
of these models.

Matter sources of the form considered here satisfy the conditions of
the generalized Birkhoff theorem. In the absence of null currents the
solutions are therefore static, whereas null charged matter generates
Vaidya-type geometries describing the time evolution of the mass and
charge.

The present work establishes a general framework for constructing exact
solutions of QTG with matter and suggests several natural extensions,
including rotating solutions and applications beyond spherical
symmetry.

\begin{acknowledgments}

This work was supported by the Natural Sciences and Engineering Research Council of Canada (NSERC). The author also gratefully acknowledges financial support from the Killam Trust.
The author is grateful to Chulmoon Yoo (Nagoya University) and Andrei Zelnikov (University of Alberta) for valuable discussions.
\end{acknowledgments}





\begin{thebibliography}{23}%
\makeatletter
\providecommand \@ifxundefined [1]{%
 \@ifx{#1\undefined}
}%
\providecommand \@ifnum [1]{%
 \ifnum #1\expandafter \@firstoftwo
 \else \expandafter \@secondoftwo
 \fi
}%
\providecommand \@ifx [1]{%
 \ifx #1\expandafter \@firstoftwo
 \else \expandafter \@secondoftwo
 \fi
}%
\providecommand \natexlab [1]{#1}%
\providecommand \enquote  [1]{``#1''}%
\providecommand \bibnamefont  [1]{#1}%
\providecommand \bibfnamefont [1]{#1}%
\providecommand \citenamefont [1]{#1}%
\providecommand \href@noop [0]{\@secondoftwo}%
\providecommand \href [0]{\begingroup \@sanitize@url \@href}%
\providecommand \@href[1]{\@@startlink{#1}\@@href}%
\providecommand \@@href[1]{\endgroup#1\@@endlink}%
\providecommand \@@startlink[1]{}%
\providecommand \@@endlink[0]{}%
\providecommand \url  [0]{\begingroup\@sanitize@url \@url }%
\providecommand \@url [1]{\endgroup\@href {#1}{\urlprefix }}%
\providecommand \urlprefix  [0]{URL }%
\providecommand \Eprint [0]{\href }%
\providecommand \doibase [0]{http://dx.doi.org/}%
\providecommand \selectlanguage [0]{\@gobble}%
\providecommand \bibinfo  [0]{\@secondoftwo}%
\providecommand \bibfield  [0]{\@secondoftwo}%
\providecommand \translation [1]{[#1]}%
\providecommand \BibitemOpen [0]{}%
\providecommand \bibitemStop [0]{}%
\providecommand \bibitemNoStop [0]{.\EOS\space}%
\providecommand \EOS [0]{\spacefactor3000\relax}%
\providecommand \BibitemShut  [1]{\csname bibitem#1\endcsname}%
\let\auto@bib@innerbib\@empty
\bibitem [{\citenamefont {Kerr}\ and\ \citenamefont
  {Schild}(1965)}]{KerrSchild1965}%
  \BibitemOpen
  \bibfield  {author} {\bibinfo {author} {\bibfnamefont {R.~P.}\ \bibnamefont
  {Kerr}}\ and\ \bibinfo {author} {\bibfnamefont {A.}~\bibnamefont {Schild}},\
  }\bibfield  {title} {\enquote {\bibinfo {title} {Some algebraically
  degenerate solutions of einstein's gravitational field equations},}\ }in\
  \href {\doibase 10.1090/psapm/017/0216846} {\emph {\bibinfo {booktitle}
  {Proceedings of Symposia in Applied Mathematics}}},\ Vol.~\bibinfo {volume}
  {17}\ (\bibinfo  {publisher} {American Mathematical Society},\ \bibinfo
  {year} {1965})\ pp.\ \bibinfo {pages} {199--209}\BibitemShut {NoStop}%
\bibitem [{\citenamefont {Visser}(2007)}]{Visser:2007fj}%
  \BibitemOpen
  \bibfield  {author} {\bibinfo {author} {\bibfnamefont {M.}~\bibnamefont
  {Visser}},\ }\bibfield  {title} {\enquote {\bibinfo {title} {{The Kerr
  spacetime: A Brief introduction}},}\ }in\ \href@noop {} {\emph {\bibinfo
  {booktitle} {{Kerr Fest: Black Holes in Astrophysics, General Relativity and
  Quantum Gravity}}}}\ (\bibinfo {year} {2007})\ \Eprint
  {http://arxiv.org/abs/0706.0622}{arXiv:0706.0622 [gr-qc]}\BibitemShut
  {NoStop}%
\bibitem [{\citenamefont {Monteiro}\ \emph {et~al.}(2014)\citenamefont
  {Monteiro}, \citenamefont {O'Connell},\ and\ \citenamefont
  {White}}]{Monteiro_2014}%
  \BibitemOpen
  \bibfield  {author} {\bibinfo {author} {\bibfnamefont {R.}~\bibnamefont
  {Monteiro}}, \bibinfo {author} {\bibfnamefont {D.}~\bibnamefont {O'Connell}},
  \ and\ \bibinfo {author} {\bibfnamefont {C.~D.}\ \bibnamefont {White}},\
  }\bibfield  {title} {\enquote {\bibinfo {title} {Black holes and the double
  copy},}\ }\href {\doibase 10.1007/jhep12(2014)056} {\bibfield  {journal}
  {\bibinfo  {journal} {Journal of High Energy Physics}\ }\textbf {\bibinfo
  {volume} {2014}} (\bibinfo {year} {2014}),\
  10.1007/jhep12(2014)056}\BibitemShut {NoStop}%
\bibitem [{\citenamefont {Luna}\ \emph {et~al.}(2015)\citenamefont {Luna},
  \citenamefont {Monteiro}, \citenamefont {O'Connell},\ and\
  \citenamefont {White}}]{Luna_2015}%
  \BibitemOpen
  \bibfield  {author} {\bibinfo {author} {\bibfnamefont {A.}~\bibnamefont
  {Luna}}, \bibinfo {author} {\bibfnamefont {R.}~\bibnamefont {Monteiro}},
  \bibinfo {author} {\bibfnamefont {D.}~\bibnamefont
  {O'Connell}}, \ and\ \bibinfo {author} {\bibfnamefont
  {C.~D.}\ \bibnamefont {White}},\ }\bibfield  {title} {\enquote {\bibinfo
  {title} {The classical double copy for {T}aub{\textendash}{NUT} spacetime},}\
  }\href {\doibase 10.1016/j.physletb.2015.09.021} {\bibfield  {journal}
  {\bibinfo  {journal} {Physics Letters B}\ }\textbf {\bibinfo {volume}
  {750}},\ \bibinfo {pages} {272} (\bibinfo {year} {2015})}\BibitemShut
  {NoStop}%
\bibitem [{\citenamefont {White}(2018)}]{White_2018}%
  \BibitemOpen
  \bibfield  {author} {\bibinfo {author} {\bibfnamefont {C.~D.}\ \bibnamefont
  {White}},\ }\bibfield  {title} {\enquote {\bibinfo {title} {The double copy:
  gravity from gluons},}\ }\href {\doibase 10.1080/00107514.2017.1415725}
  {\bibfield  {journal} {\bibinfo  {journal} {Contemporary Physics}\ }\textbf
  {\bibinfo {volume} {59}},\ \bibinfo {pages} {109} (\bibinfo {year}
  {2018})}\BibitemShut {NoStop}%
\bibitem [{\citenamefont {Bern}\ \emph {et~al.}(2019)\citenamefont {Bern},
  \citenamefont {Carrasco}, \citenamefont {Chiodaroli}, \citenamefont
  {Johansson},\ and\ \citenamefont {Roiban}}]{bern2019duality}%
  \BibitemOpen
  \bibfield  {author} {\bibinfo {author} {\bibfnamefont {Z.}~\bibnamefont
  {Bern}}, \bibinfo {author} {\bibfnamefont {J.~J.}\ \bibnamefont {Carrasco}},
  \bibinfo {author} {\bibfnamefont {M.}~\bibnamefont {Chiodaroli}}, \bibinfo
  {author} {\bibfnamefont {H.}~\bibnamefont {Johansson}}, \ and\ \bibinfo
  {author} {\bibfnamefont {R.}~\bibnamefont {Roiban}},\ }\href@noop {}
  {\enquote {\bibinfo {title} {The duality between color and kinematics and its
  applications},}\ } (\bibinfo {year} {2019}),\ \Eprint
  {http://arxiv.org/abs/1909.01358}{arXiv:1909.01358 [hep-th]}\BibitemShut
  {NoStop}%
\bibitem [{\citenamefont {Frolov}(2026)}]{Frolov:2025ddw}%
  \BibitemOpen
  \bibfield  {author} {\bibinfo {author} {\bibfnamefont {V.~P.}\ \bibnamefont
  {Frolov}},\ }\bibfield  {title} {\enquote {\bibinfo {title}
  {{Quasitopological gravity and double-copy formalism}},}\ }\href {\doibase
  10.1103/mbdj-tqtv} {\bibfield  {journal} {\bibinfo  {journal} {Phys. Rev. D}\
  }\textbf {\bibinfo {volume} {113}},\ \bibinfo {pages} {064023} (\bibinfo
  {year} {2026})},\ \Eprint {http://arxiv.org/abs/2512.14674}{arXiv:2512.14674
  [gr-qc]}\BibitemShut {NoStop}%
\bibitem [{\citenamefont {Bueno}\ \emph {et~al.}(2020)\citenamefont {Bueno},
  \citenamefont {Cano},\ and\ \citenamefont {Hennigar}}]{Bueno:2019ycr}%
  \BibitemOpen
  \bibfield  {author} {\bibinfo {author} {\bibfnamefont {P.}~\bibnamefont
  {Bueno}}, \bibinfo {author} {\bibfnamefont {P.~A.}\ \bibnamefont {Cano}}, \
  and\ \bibinfo {author} {\bibfnamefont {R.~A.}\ \bibnamefont {Hennigar}},\
  }\bibfield  {title} {\enquote {\bibinfo {title} {{(Generalized)
  quasi-topological gravities at all orders}},}\ }\href {\doibase
  10.1088/1361-6382/ab5410} {\bibfield  {journal} {\bibinfo  {journal} {Class.
  Quant. Grav.}\ }\textbf {\bibinfo {volume} {37}},\ \bibinfo {pages} {015002}
  (\bibinfo {year} {2020})},\ \Eprint
  {http://arxiv.org/abs/1909.07983}{arXiv:1909.07983 [hep-th]}\BibitemShut
  {NoStop}%
\bibitem [{\citenamefont {Bueno}\ \emph {et~al.}(2019)\citenamefont {Bueno},
  \citenamefont {Cano}, \citenamefont {Moreno},\ and\ \citenamefont
  {Murcia}}]{Bueno:2019ltp}%
  \BibitemOpen
  \bibfield  {author} {\bibinfo {author} {\bibfnamefont {P.}~\bibnamefont
  {Bueno}}, \bibinfo {author} {\bibfnamefont {P.~A.}\ \bibnamefont {Cano}},
  \bibinfo {author} {\bibfnamefont {J.}~\bibnamefont {Moreno}}, \ and\ \bibinfo
  {author} {\bibfnamefont {{\'A}.}~\bibnamefont {Murcia}},\ }\bibfield  {title}
  {\enquote {\bibinfo {title} {{All higher-curvature gravities as Generalized
  quasi-topological gravities}},}\ }\href {\doibase 10.1007/JHEP11(2019)062}
  {\bibfield  {journal} {\bibinfo  {journal} {JHEP}\ }\textbf {\bibinfo
  {volume} {11}},\ \bibinfo {pages} {062} (\bibinfo {year} {2019})},\ \Eprint
  {http://arxiv.org/abs/1906.00987}{arXiv:1906.00987 [hep-th]}\BibitemShut
  {NoStop}%
\bibitem [{\citenamefont {Bueno}\ \emph
  {et~al.}(2025{\natexlab{a}})\citenamefont {Bueno}, \citenamefont {Cano},\
  and\ \citenamefont {Hennigar}}]{Bueno:2024dgm}%
  \BibitemOpen
  \bibfield  {author} {\bibinfo {author} {\bibfnamefont {P.}~\bibnamefont
  {Bueno}}, \bibinfo {author} {\bibfnamefont {P.~A.}\ \bibnamefont {Cano}}, \
  and\ \bibinfo {author} {\bibfnamefont {R.~A.}\ \bibnamefont {Hennigar}},\
  }\bibfield  {title} {\enquote {\bibinfo {title} {Regular black holes from
  pure gravity},}\ }\href {\doibase 10.1016/j.physletb.2025.139260} {\bibfield
  {journal} {\bibinfo  {journal} {Phys. Lett. B}\ }\textbf {\bibinfo {volume}
  {861}},\ \bibinfo {pages} {139260} (\bibinfo {year} {2025}{\natexlab{a}})},\
  \Eprint {http://arxiv.org/abs/2403.04827}{arXiv:2403.04827
  [gr-qc]}\BibitemShut {NoStop}%
\bibitem [{\citenamefont {Frolov}\ \emph {et~al.}(2025)\citenamefont {Frolov},
  \citenamefont {Koek}, \citenamefont {Pinedo~Soto},\ and\ \citenamefont
  {Zelnikov}}]{Frolov:2024lza}%
  \BibitemOpen
  \bibfield  {author} {\bibinfo {author} {\bibfnamefont {V.~P.}\ \bibnamefont
  {Frolov}}, \bibinfo {author} {\bibfnamefont {A.}~\bibnamefont {Koek}},
  \bibinfo {author} {\bibfnamefont {J.}~\bibnamefont {Pinedo~Soto}}, \ and\
  \bibinfo {author} {\bibfnamefont {A.}~\bibnamefont {Zelnikov}},\ }\bibfield
  {title} {\enquote {\bibinfo {title} {Regular black holes inspired by
  quasitopological gravity},}\ }\href {\doibase 10.1103/PhysRevD.111.044034}
  {\bibfield  {journal} {\bibinfo  {journal} {Phys. Rev. D}\ }\textbf {\bibinfo
  {volume} {111}},\ \bibinfo {pages} {044034} (\bibinfo {year} {2025})},\
  \Eprint {http://arxiv.org/abs/2411.16050}{arXiv:2411.16050
  [gr-qc]}\BibitemShut {NoStop}%
\bibitem [{\citenamefont {Bueno}\ \emph
  {et~al.}(2025{\natexlab{b}})\citenamefont {Bueno}, \citenamefont {Cano},
  \citenamefont {Hennigar},\ and\ \citenamefont {Murcia}}]{Bueno:2025collapse}%
  \BibitemOpen
  \bibfield  {author} {\bibinfo {author} {\bibfnamefont {P.}~\bibnamefont
  {Bueno}}, \bibinfo {author} {\bibfnamefont {P.~A.}\ \bibnamefont {Cano}},
  \bibinfo {author} {\bibfnamefont {R.~A.}\ \bibnamefont {Hennigar}}, \ and\
  \bibinfo {author} {\bibfnamefont {A.~J.}\ \bibnamefont {Murcia}},\ }\bibfield
   {title} {\enquote {\bibinfo {title} {Regular black holes from thin-shell
  collapse},}\ }\href {\doibase 10.1103/PhysRevD.111.104009} {\bibfield
  {journal} {\bibinfo  {journal} {Phys. Rev. D}\ }\textbf {\bibinfo {volume}
  {111}},\ \bibinfo {pages} {104009} (\bibinfo {year}
  {2025}{\natexlab{b}})}\BibitemShut {NoStop}%
\bibitem [{\citenamefont {Bueno}\ \emph
  {et~al.}(2025{\natexlab{c}})\citenamefont {Bueno}, \citenamefont {Cano},
  \citenamefont {Hennigar},\ and\ \citenamefont
  {Murcia}}]{Bueno:2025formation}%
  \BibitemOpen
  \bibfield  {author} {\bibinfo {author} {\bibfnamefont {P.}~\bibnamefont
  {Bueno}}, \bibinfo {author} {\bibfnamefont {P.~A.}\ \bibnamefont {Cano}},
  \bibinfo {author} {\bibfnamefont {R.~A.}\ \bibnamefont {Hennigar}}, \ and\
  \bibinfo {author} {\bibfnamefont {A.~J.}\ \bibnamefont {Murcia}},\ }\bibfield
   {title} {\enquote {\bibinfo {title} {Dynamical formation of regular black
  holes},}\ }\href {\doibase 10.1103/PhysRevLett.134.181401} {\bibfield
  {journal} {\bibinfo  {journal} {Phys. Rev. Lett.}\ }\textbf {\bibinfo
  {volume} {134}},\ \bibinfo {pages} {181401} (\bibinfo {year}
  {2025}{\natexlab{c}})}\BibitemShut {NoStop}%
\bibitem [{\citenamefont {Bueno}\ \emph
  {et~al.}(2026{\natexlab{a}})\citenamefont {Bueno}, \citenamefont {Hennigar},
  \citenamefont {Murcia},\ and\ \citenamefont
  {Vicente-Cano}}]{Bueno:2026singularmatter}%
  \BibitemOpen
  \bibfield  {author} {\bibinfo {author} {\bibfnamefont {P.}~\bibnamefont
  {Bueno}}, \bibinfo {author} {\bibfnamefont {R.~A.}\ \bibnamefont {Hennigar}},
  \bibinfo {author} {\bibfnamefont {A.~J.}\ \bibnamefont {Murcia}}, \ and\
  \bibinfo {author} {\bibfnamefont {A.}~\bibnamefont {Vicente-Cano}},\
  }\bibfield  {title} {\enquote {\bibinfo {title} {Regular geometries from
  singular matter in quasitopological gravity},}\ }\href {\doibase
  10.1103/3mk3-j8vs} {\bibfield  {journal} {\bibinfo  {journal} {Phys. Rev. D}\
  }\textbf {\bibinfo {volume} {114}},\ \bibinfo {pages} {024070} (\bibinfo
  {year} {2026}{\natexlab{a}})}\BibitemShut {NoStop}%
\bibitem [{\citenamefont {Frolov}\ and\ \citenamefont
  {Zelnikov}(2026)}]{Frolov:2026massinflation}%
  \BibitemOpen
  \bibfield  {author} {\bibinfo {author} {\bibfnamefont {V.~P.}\ \bibnamefont
  {Frolov}}\ and\ \bibinfo {author} {\bibfnamefont {A.}~\bibnamefont
  {Zelnikov}},\ }\bibfield  {title} {\enquote {\bibinfo {title} {Regular black
  holes in quasitopological gravity: Null shells and mass inflation},}\
  }\href@noop {} {\  (\bibinfo {year} {2026})},\ \Eprint
  {http://arxiv.org/abs/2601.01861}{arXiv:2601.01861 [gr-qc]}\BibitemShut
  {NoStop}%
\bibitem [{\citenamefont {Bueno}\ \emph
  {et~al.}(2026{\natexlab{b}})\citenamefont {Bueno}, \citenamefont {Cano},
  \citenamefont {Hennigar},\ and\ \citenamefont {Murcia}}]{Bueno:2026nonlocal}%
  \BibitemOpen
  \bibfield  {author} {\bibinfo {author} {\bibfnamefont {P.}~\bibnamefont
  {Bueno}}, \bibinfo {author} {\bibfnamefont {P.~A.}\ \bibnamefont {Cano}},
  \bibinfo {author} {\bibfnamefont {R.~A.}\ \bibnamefont {Hennigar}}, \ and\
  \bibinfo {author} {\bibfnamefont {A.~J.}\ \bibnamefont {Murcia}},\ }\bibfield
   {title} {\enquote {\bibinfo {title} {Regular black holes in nonlocal
  quasitopological gravity},}\ }\href@noop {} {\  (\bibinfo {year}
  {2026}{\natexlab{b}})},\ \Eprint
  {http://arxiv.org/abs/2607.07790}{arXiv:2607.07790 [gr-qc]}\BibitemShut
  {NoStop}%
\bibitem [{\citenamefont {Borissova}(2026{\natexlab{a}})}]{Borissova:2026rbi}%
  \BibitemOpen
  \bibfield  {author} {\bibinfo {author} {\bibfnamefont {J.}~\bibnamefont
  {Borissova}},\ }\bibfield  {title} {\enquote {\bibinfo {title}
  {{$g_{tt}g_{rr} =-1$ black hole thermodynamics in extended quasi-topological
  gravity}},}\ }\href@noop {} {\  (\bibinfo {year} {2026}{\natexlab{a}})},\
  \Eprint {http://arxiv.org/abs/2604.24101}{arXiv:2604.24101
  [gr-qc]}\BibitemShut {NoStop}%
\bibitem [{\citenamefont {Borissova}(2026{\natexlab{b}})}]{Borissova:2026prd}%
  \BibitemOpen
  \bibfield  {author} {\bibinfo {author} {\bibfnamefont {J.}~\bibnamefont
  {Borissova}},\ }\bibfield  {title} {\enquote {\bibinfo {title} {{Formation of
  extremal regular black holes}},}\ }\href@noop {} {\  (\bibinfo {year}
  {2026}{\natexlab{b}})},\ \Eprint
  {http://arxiv.org/abs/2608.02908}{arXiv:2608.02908 [gr-qc]}\BibitemShut
  {NoStop}%
\bibitem [{\citenamefont {Borissova}\ and\ \citenamefont
  {Carballo-Rubio}(2026)}]{Borissova:2026wmn}%
  \BibitemOpen
  \bibfield  {author} {\bibinfo {author} {\bibfnamefont {J.}~\bibnamefont
  {Borissova}}\ and\ \bibinfo {author} {\bibfnamefont {R.}~\bibnamefont
  {Carballo-Rubio}},\ }\bibfield  {title} {\enquote {\bibinfo {title} {{Regular
  black holes from pure gravity in four dimensions}},}\ }\href {\doibase
  10.1103/6x2z-qbkh} {\bibfield  {journal} {\bibinfo  {journal} {Phys. Rev. D}\
  }\textbf {\bibinfo {volume} {113}},\ \bibinfo {pages} {124004} (\bibinfo
  {year} {2026})},\ \Eprint {http://arxiv.org/abs/2602.16773}{arXiv:2602.16773
  [gr-qc]}\BibitemShut {NoStop}%
\bibitem [{\citenamefont {Frolov}\ \emph {et~al.}(2026)\citenamefont {Frolov},
  \citenamefont {Yoo},\ and\ \citenamefont {Zelnikov}}]{Frolov:2026tft}%
  \BibitemOpen
  \bibfield  {author} {\bibinfo {author} {\bibfnamefont {V.~P.}\ \bibnamefont
  {Frolov}}, \bibinfo {author} {\bibfnamefont {C.-M.}\ \bibnamefont {Yoo}}, \
  and\ \bibinfo {author} {\bibfnamefont {A.}~\bibnamefont {Zelnikov}},\
  }\bibfield  {title} {\enquote {\bibinfo {title} {{Vaidya-Type Solutions of
  Quasitopological Gravity Interacting with Nonlinear Electrodynamics}},}\
  }\href@noop {} {\  (\bibinfo {year} {2026})},\ \Eprint
  {http://arxiv.org/abs/2607.22856}{arXiv:2607.22856 [gr-qc]}\BibitemShut
  {NoStop}%
\bibitem [{\citenamefont {Hennigar}\ \emph {et~al.}(2025)\citenamefont
  {Hennigar}, \citenamefont {Kubiz{\v{n}}{\'a}k}, \citenamefont {Murk},\ and\
  \citenamefont {Soranidis}}]{Hennigar:2025yqm}%
  \BibitemOpen
  \bibfield  {author} {\bibinfo {author} {\bibfnamefont {R.~A.}\ \bibnamefont
  {Hennigar}}, \bibinfo {author} {\bibfnamefont {D.}~\bibnamefont
  {Kubiz{\v{n}}{\'a}k}}, \bibinfo {author} {\bibfnamefont {S.}~\bibnamefont
  {Murk}}, \ and\ \bibinfo {author} {\bibfnamefont {I.}~\bibnamefont
  {Soranidis}},\ }\bibfield  {title} {\enquote {\bibinfo {title}
  {{Thermodynamics of regular black holes in anti-de Sitter space}},}\ }\href
  {\doibase 10.1007/JHEP11(2025)121} {\bibfield  {journal} {\bibinfo  {journal}
  {JHEP}\ }\textbf {\bibinfo {volume} {11}},\ \bibinfo {pages} {121} (\bibinfo
  {year} {2025})},\ \Eprint {http://arxiv.org/abs/2505.11623}{arXiv:2505.11623
  [gr-qc]}\BibitemShut {NoStop}%
\bibitem [{\citenamefont {Hao}\ \emph {et~al.}(2026)\citenamefont {Hao},
  \citenamefont {Jing},\ and\ \citenamefont {Wang}}]{Hao:2025utc}%
  \BibitemOpen
  \bibfield  {author} {\bibinfo {author} {\bibfnamefont {C.-H.}\ \bibnamefont
  {Hao}}, \bibinfo {author} {\bibfnamefont {J.}~\bibnamefont {Jing}}, \ and\
  \bibinfo {author} {\bibfnamefont {J.}~\bibnamefont {Wang}},\ }\bibfield
  {title} {\enquote {\bibinfo {title} {{Charged regular black holes from
  quasi-topological gravities in D {\ensuremath{\geq}} 5}},}\ }\href {\doibase
  10.1088/1475-7516/2026/05/067} {\bibfield  {journal} {\bibinfo  {journal}
  {JCAP}\ }\textbf {\bibinfo {volume} {05}},\ \bibinfo {pages} {067} (\bibinfo
  {year} {2026})},\ \Eprint {http://arxiv.org/abs/2512.04604}{arXiv:2512.04604
  [gr-qc]}\BibitemShut {NoStop}%
\bibitem [{\citenamefont {Pinedo~Soto}\ and\ \citenamefont
  {Frolov}(2026)}]{PinedoSoto:2026hfm}%
  \BibitemOpen
  \bibfield  {author} {\bibinfo {author} {\bibfnamefont {J.}~\bibnamefont
  {Pinedo~Soto}}\ and\ \bibinfo {author} {\bibfnamefont {V.~P.}\ \bibnamefont
  {Frolov}},\ }\bibfield  {title} {\enquote {\bibinfo {title} {Charged black
  holes in quasitopological gravity coupled to {B}orn-{I}nfeld nonlinear
  electrodynamics},}\ }\href {\doibase 10.1103/8px4-2pyk} {\bibfield  {journal}
  {\bibinfo  {journal} {Phys. Rev. D}\ }\textbf {\bibinfo {volume} {113}},\
  \bibinfo {pages} {124044} (\bibinfo {year} {2026})},\ \Eprint
  {http://arxiv.org/abs/2604.06632}{arXiv:2604.06632 [gr-qc]}\BibitemShut
  {NoStop}%
\end{thebibliography}

%

\end{document}